\documentclass[aps,twocolumn,preprintnumbers,nofootinbib,10pt]{revtex4-1}%

\pdfoutput=1
\usepackage{amsmath,amsthm,amssymb,multirow,psfrag}
\usepackage{epsfig}
\usepackage{color}

\graphicspath{{./Figures/}}

\begin{document}

\def\lsim{\mathrel{\rlap{\lower4pt\hbox{\hskip1pt$\sim$}}
  \raise1pt\hbox{$<$}}}
\def\gsim{\mathrel{\rlap{\lower4pt\hbox{\hskip1pt$\sim$}}
  \raise1pt\hbox{$>$}}}
\newcommand{\vev}[1]{ \left\langle {#1} \right\rangle }
\newcommand{\bra}[1]{ \langle {#1} | }
\newcommand{\ket}[1]{ | {#1} \rangle }
\newcommand{\ev}{ {\rm eV} }
\newcommand{\kev}{{\rm keV}}
\newcommand{\mev}{{SM\rm MeV}}
\newcommand{\tev}{{\rm TeV}}
\newcommand{\mpl}{$M_{Pl}$}
\newcommand{\mw}{$M_{W}$}
\newcommand{\Ft}{F_{T}}
\newcommand{\Zparity}{\mathbb{Z}_2}
\newcommand{\BLambda}{\boldsymbol{\lambda}}
\newcommand{\met}{\;\not\!\!\!{E}_T}

\newcommand{\beq}{\begin{equation}}
\newcommand{\eeq}{\end{equation}}
\newcommand{\bea}{\begin{eqnarray}}
\newcommand{\eea}{\end{eqnarray}}
\newcommand{\nn}{\nonumber}
\newcommand{\gev}{{\mathrm GeV}}
\newcommand{\hc}{\mathrm{h.c.}}
\newcommand{\eps}{\epsilon}

\newcommand{\cO}{{\cal O}}
\newcommand{\cL}{{\cal L}}
\newcommand{\cM}{{\cal M}}

\graphicspath{{./Figures/}}

\newcommand{\fref}[1]{Fig.~\ref{fig:#1}} 
\newcommand{\eref}[1]{Eq.~\eqref{eq:#1}} 
\newcommand{\aref}[1]{Appendix~\ref{app:#1}}
\newcommand{\sref}[1]{Sec.~\ref{sec:#1}}
\newcommand{\tref}[1]{Table~\ref{tab:#1}}  

\def\TY#1{{\bf  \textcolor{red}{[TY: {#1}]}}}
\newcommand{\draftnote}[1]{{\bf\color{blue} #1}}
\newcommand{\draftnoteR}[1]{{\bf\color{red} #1}}

\title{ {\bf \Large{Diphoton and Diboson Probes of\\ Fermiophobic Higgs Bosons at the LHC}\normalsize}}
\author{\bf{Antonio Delgado$^{a}$,~Mateo Garcia-Pepin$^{b}$,~Mariano Quir\'{o}s$^{b,c}$,\\~Jos\'{e} Santiago$^{d}$,~Roberto Vega-Morales$^{d}$}}

\vspace{-.1cm}
\affiliation{
$^{a}$Department of Physics, University of Notre Dame, Notre Dame, IN 46556, USA\\
$^{b}$Institut de F\'{i}sica d'Altes Energies (IFAE), The Barcelona Institute of  Science and Technology (BIST)\\ Universitat Aut$\grave{o}$noma de Barcelona, Barcelona, Spain\\
$^{c}$Instituci\'{o} Catalana de Recerca i Estudis Avan\c{c}ats (ICREA), Barcelona, Spain\\
$^{d}$Departamento de F\'{i}sica Te\'{o}rica y del Cosmos and CAFPE,
Universidad de Granada, Campus de Fuentenueva, E-18071 Granada, Spain
}

\begin{abstract}

Extensions of the Standard Model Higgs sector with electroweak charged scalars can possess exotic `Higgs' bosons with vanishing or suppressed couplings to Standard Model fermions.~These `fermiophobic' scalars, which cannot be produced via gluon fusion, are constrained by LHC measurements of the 125~GeV Higgs boson to have a small vacuum expectation value.~This implies that vector boson fusion and associated vector boson production are in general suppressed rendering conventional Higgs searches insensitive.~However, Drell-Yan Higgs pair production, which is not present in the SM, can be sizeable even in the limit of vanishing exotic Higgs vacuum expectation value.~We utilize this to show that diphoton searches at 8 TeV LHC already rule out a large class of neutral fermiophobic Higgs bosons below $\sim 110$~GeV.~This includes fermiophobic scalars found in two Higgs doublet as well as Higgs triplet and Georgi-Machacek type models.~Our results extend the only relevant limit on fermiophobic Higgs bosons obtained by a recent CDF analysis of $4\gamma + X$ Tevatron data.~Furthermore, diphoton limits are independent of the decay of the second Higgs boson and thus apply even for degenerate masses in contrast to the CDF search.~We also find that if the fermiophobic Higgs has very enhanced couplings to photons, masses as large as $\sim 150$~GeV can be ruled out while if these couplings are somehow highly suppressed, masses below $\sim 90$~GeV can still be ruled out.~Finally, we show that $WW$ and $ZZ$ diboson searches may serve as complementary probes for masses above the diphoton limit up to $\sim 250$~GeV and discuss prospects at 13 TeV LHC.
\end{abstract}

\preprint{UG-FT 320/16}
\preprint{CAFPE 190/16}

\maketitle

\section{Introduction} \label{sec:intro} 

The discovery of the Higgs boson~\cite{:2012gu,:2012gk} has provided the first direct window into the mechanism of electroweak symmetry breaking (EWSB).~Many models of EWSB and extensions of the Standard Model (SM) predict enlarged Higgs sectors, in which the spectrum includes extra electroweak charged scalars beyond the SM Higgs doublet.~In a number of models, neutral Higgs bosons with suppressed or vanishing couplings to SM fermions are present at the weak scale after EWSB.~These `fermiophobic' Higgs bosons have many generic phenomenological features which have been considered for some time~\cite{Pois:1993ay,Stange:1993ya,Diaz:1994pk,Akeroyd:1995hg,Akeroyd:1998ui,Barroso:1999bf,Brucher:1999tx,Landsberg:2000ht,Akeroyd:2003bt,Akeroyd:2003jp,Akeroyd:2003xi,Akeroyd:2005pr,Akeroyd:2010eg} and searched for previously at LEP~\cite{Abreu:2001ib,Abbiendi:2002yc,Heister:2002ub,Achard:2002jh}, Tevatron~\cite{Abbott:1998vv,Affolder:2001hx}, and LHC~\cite{Chatrchyan:2013sfs}.~They can be found in `Type I' two Higgs doublet models~\cite{Haber:1978jt} in the large $\tan\beta$ limit~\cite{Mrenna:2000qh,Akeroyd:2010eg,Gabrielli:2012hd} as well as Higgs triplet models~\cite{Blank:1997qa,Chang:2003zn,Chen:2008jg,Delgado:2012sm,Delgado:2013zfa,deBlas:2013epa,Bandyopadhyay:2014tha,Bandyopadhyay:2014raa,Alvarado:2015yna} including the well known Georgi-Macachek (GM) model~\cite{Georgi:1985nv} and its variations~\cite{Chanowitz:1985ug,Gunion:1989ci,Gunion:1990dt,Gunion:1989ci,Gunion:1990dt,Aoki:2007ah,Chiang:2012cn,Chiang:2013rua,Hartling:2014zca,Logan:2015xpa} or their supersymmetric incarnations~\cite{Cort:2013foa,Garcia-Pepin:2014yfa,Delgado:2015aha,Delgado:2015bwa,Delgado:2015hmy,Garcia-Pepin:2016hvs}.~They can also appear in non-minimal composite Higgs models~\cite{Mrazek:2011iu,Bellazzini:2014yua}.

Since these fermiophobic Higgs scalars do not couple to quarks, production via gluon fusion is not available.~Furthermore, as pointed out in~\cite{Akeroyd:2003xi,Akeroyd:2005pr}, if the vacuum expectation value (vev) of the fermiophobic Higgs is small, vector boson fusion (VBF) and associated Higgs vector boson production (VH) quickly become highly suppressed.~Since these are the dominant production mechanisms in the SM, they have been assumed as the production mechanisms in almost all Higgs boson searches regardless of if they are fermiophobic or not.~On the other hand since LHC measurements of the 125~GeV Higgs boson couplings~\cite{Khachatryan:2014kca} seem to indicate a SM-like Higgs boson~\cite{Falkowski:2013dza}, this implies a small vev for any additional exotic Higgs boson.~As these measurements increase in precision without observing a deviation from the SM prediction, previous collider searches for fermiophobic Higgs bosons, which assumed SM-like production mechanisms, become increasingly obsolete.

However, Drell-Yan (DY) Higgs pair production, which is not present in the SM, can be sizable even in the limit of small exotic Higgs vev~\cite{Akeroyd:2003xi,Akeroyd:2005pr}.~Furthermore, as pointed out many times~\cite{Mrenna:2000qh,Landsberg:2000ht,Arhrib:2003ph,Akeroyd:2007yh,Akeroyd:2012ms}, since there is no $b\bar{b}$ decay to compete with, neutral fermiophobic Higgs scalars (which we refer to as $H_F^0$)  at low masses can have large branching ratios to vector boson pairs and in particular photons.~This can be combined with DY pair production to place stringent constraints on light fermiophobic Higgs bosons using multiphoton final states.~In particular, the $W$ boson mediated $H^{\pm}H_F^0$ production channel (see~\fref{HHprod}), followed by $H^{\pm} \to W^{\pm}H^{0}_F$ and $H^{0}_F \to \gamma\gamma$ decays, leads to a $4\gamma + X$ final state, which has been proposed as a probe~\cite{Akeroyd:2003xi,Akeroyd:2005pr} of fermiophobic Higgs bosons at high energy colliders.~Clearly the $H^{\pm} \to W^{\pm}H^{0}_F$ decay requires a mass splitting between the charged and neutral Higgs and, in particular, $M_{H^{\pm}} > M_{H^{0}_F}$.

The lone experimental search to utilize this DY pair production to multi-photon channel to search for a fermiophobic Higgs is a very recent CDF analysis of previously collected Tevatron data~\cite{Aaltonen:2016fnw}.~This was applied to fermiophobic Higgs bosons found in Type I two Higgs doublet models to put constraints for the first time and, in particular, rule out a neutral fermiophobic Higgs boson below $100$~GeV.~Constraints in the two dimensional plane of the charged and neutral Higgs boson masses were also obtained.~Of course in the limit where the mass splitting goes to zero this multiphoton search can be evaded.~In models with custodial symmetry~\cite{Sikivie:1980hm} in the Higgs sector, which are motivated by electroweak precision data, degenerate masses between the neutral and charged Higgs is commonly found (at tree level).~This makes CDF searches in the $4\gamma + X$ channel insensitive to these fermiophobic custodial Higgs scalars~\footnote{Of course if there are additional Higgs scalars which are in different custodial representations than $H_F^0$, additional Higgs pair production mechanisms with non-degenerate masses can become available allowing for $4\gamma + X$ limits to again be applied.}.~Clearly these searches are also insensitive when $M_{H^{\pm}} < M_{H^{0}}$

In this paper we emphasize that the $W$ mediated $H^{\pm}H_F^0$ pair production can also be combined with conventional \emph{diphoton} searches to probe neutral fermiophobic Higgs bosons.~While the signal to background ratio is worse than in $4\gamma + X$~\cite{Aaltonen:2016fnw}, diphoton searches have the advantage that, being more inclusive, are more model independent and can probe neutral fermiophobic Higgs bosons without any reference to the second Higgs boson decay.~In particular, they can be applied even in the custodial limit of degenerate masses as well as when $M_{H^{\pm}} < M_{H^{0}}$ or if the charged Higgs decays in a way that is difficult to observe.~We find that while Tevatron diphoton searches are not sensitive to fermiophobic Higgs bosons, the larger production cross sections at 8 TeV LHC allow a neutral fermiophobic Higgs boson below $90 - 150$~GeV to be ruled out depending on particular model assumptions.~For similar assumptions, we find that stronger bounds than those obtained in $4\gamma + X$ searches at Tevatron can be obtained with 8 TeV diphoton searches at LHC.~We also examine, combining the Higgs pair production with $WW$ and $ZZ$ diboson searches as a complementary probe to diphoton searches, for larger fermiophobic Higgs masses up to $\sim 250$~GeV.

Finally, we pay particular attention to the specific case of a fermiophobic custodial fiveplet scalar found in all incarnations of custodial Higgs triplet models~\cite{Georgi:1985nv,Hartling:2014zca,Cort:2013foa} in which the neutral and charged Higgs scalars are predicted to be degenerate.~Thus the CDF $4\gamma + X$ search~\cite{Aaltonen:2016fnw} cannot be applied to this case.~We show for the first time that when the $W$ boson loop (see~\fref{HtoVA}) dominates the effective couplings to photons, a custodial fiveplet scalar below $\sim 110$~GeV is ruled out by 8 TeV LHC diphoton searches independently of the Higgs triplet vev.~Larger masses possibly up to $\sim 150$~GeV can also be ruled out if charged scalar loops produce large constructive contributions to the effective photon couplings.~We also find that diboson searches, and in particular $ZZ$ searches, may be useful for higher masses allowing us to potentially obtain limits again  for custodial fiveplet masses up to $\sim 250$~GeV independently of the Higgs triplet vev.

The paper is organized as follows:~in~\sref{proddec} we review the relevant aspects of fermiophobic production and decay for LHC diboson searches.~In~\sref{LHC} we examine diphoton and diboson searches at 8 TeV LHC for generic fermiophobic Higgs scenearios.~Finally in~\sref{H5} we examine the particular case of a custodial fiveplet scalar before summarizing our conclusions in~\sref{sum}.

\section{Fermiophobic Higgs Boson \\
~~~~~~Production and Decay}\label{sec:proddec}

Here we review production and decay of fermiophobic Higgs bosons focusing on the aspects most relevant for LHC diphoton and diboson searches.~In particular we focus on the limit of small exotic Higgs vev in which the DY Higgs pair production mechanism is dominant.~A more detailed discussion of fermiophobic Higgs production and decays can be found in~\cite{Stange:1993ya,Diaz:1994pk,Akeroyd:1995hg,Akeroyd:1998ui,Barroso:1999bf,Landsberg:2000ht,Akeroyd:2003bt,Akeroyd:2003jp,Akeroyd:2003xi,Akeroyd:2005pr}, and references therein, to which we refer the reader for details.

\subsection{Higgs Pair Production}

Any extension of the SM Higgs sector by electroweak charged scalars will
possess the pair production channel mediated by a $W$ boson shown
in~\fref{HHprod}.~Here we take $H_F^0$ to generically represent our neutral fermiophobic Higgs boson and assume it to be CP even while $H_N^{\pm}$ is in an arbitrary $SU(2)_L \otimes U(1)_Y$ representation labeled by $N$ which may (or may not) be the same representation which $H_F^0$ belongs to.~The corresponding diagram involving a $Z$ boson can arise when $H_N^{\pm}$ is replaced by a neutral CP odd scalar, but is subdominant to the $W$ mediated channel~\cite{Akeroyd:2003xi}.~In general the neutral and charged components in~\fref{HHprod} can have different masses, but as long as the mass splitting is not too large and both are sufficiently light to be produced on-shell, it will not qualitatively affect our discussion since we will only be concerned with the neutral fermiophobic Higgs decay.
\begin{figure}[tbh]
\begin{center}
\includegraphics[scale=.6]{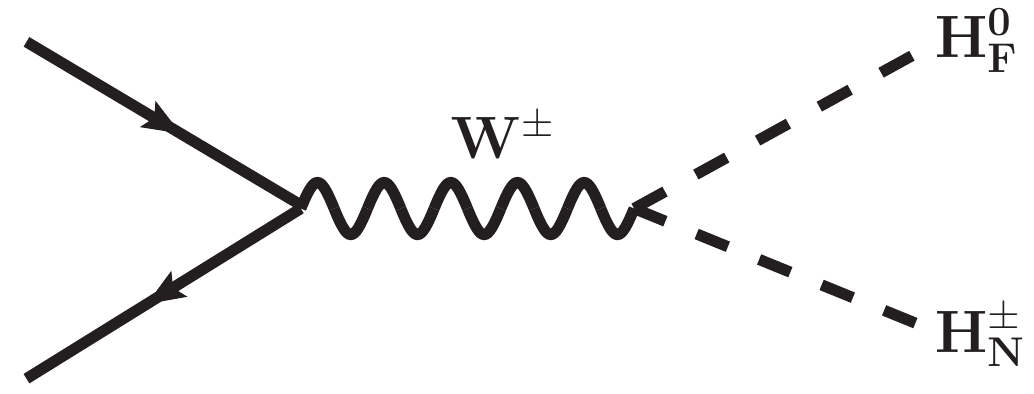}
\end{center}
\caption{The dominant contribution to Higgs pair production in extensions of the Standard Model Higgs sector.}
\label{fig:HHprod}
\end{figure}

We can write the $WHH$ vertex schematically as,
\bea\label{eq:gwhh}
V_{WHH} \equiv i g C_N (p_1 - p_2)^\mu
\eea
where $C_N$ is fixed by the $SU(2)_L$ representation and
$p_{1}, p_{2}$ are the four momenta of the incoming and outgoing
scalar momenta.~Once EWSB occurs there may (or may not) be a dependence on the exotic Higgs vev and mass mixing angles introduced into the vertex, depending on if the gauge and mass eigenstates are `aligned'.~The key point is that, unlike the terms which generate the $W$ and $Z$ masses or the single coupling of $H$ to pairs of electroweak vector bosons, $V_{WHH}$ does not \emph{necessarily} depend on the exotic Higgs vev and, more importantly, does not go to zero in the limit of vanishing vev.~A more detailed discussion of how the vertex in~\eref{gwhh} can depend on these various mixing angles in the context of the two Higgs doublet or Higgs triplet models can be found in~\cite{Georgi:1985nv,Akeroyd:2003bt,Akeroyd:2010eg,Cort:2013foa,Hartling:2014zca} and references therein.

To see roughly how large these Higgs pair production cross sections are, we show in~\fref{Hprod} leading order cross sections for various channels (thick solid curves) involving $H_F^0$ at the LHC with $\sqrt{s}=8$~TeV in the mass range $45 - 250$~GeV.~In these curves we have factored out any group theory factors or mass mixing angles which could enter in the vertex in~\eref{gwhh} so that the coefficient is simply given by the $SU(2)$ gauge coupling $g$.~The curves for any particular model can be obtained by trivial rescaling with $(C_N)^2$ and will not qualitatively change this discussion which is largely for intuition purposes.~Our results are obtained from Madgraph~\cite{Alwall:2014hca} using a modified version of the GM model implementation of~\cite{Hartling:2014xma} and rescaling appropriately.
\begin{figure}[tbh]
\begin{center}
\includegraphics[scale=.4]{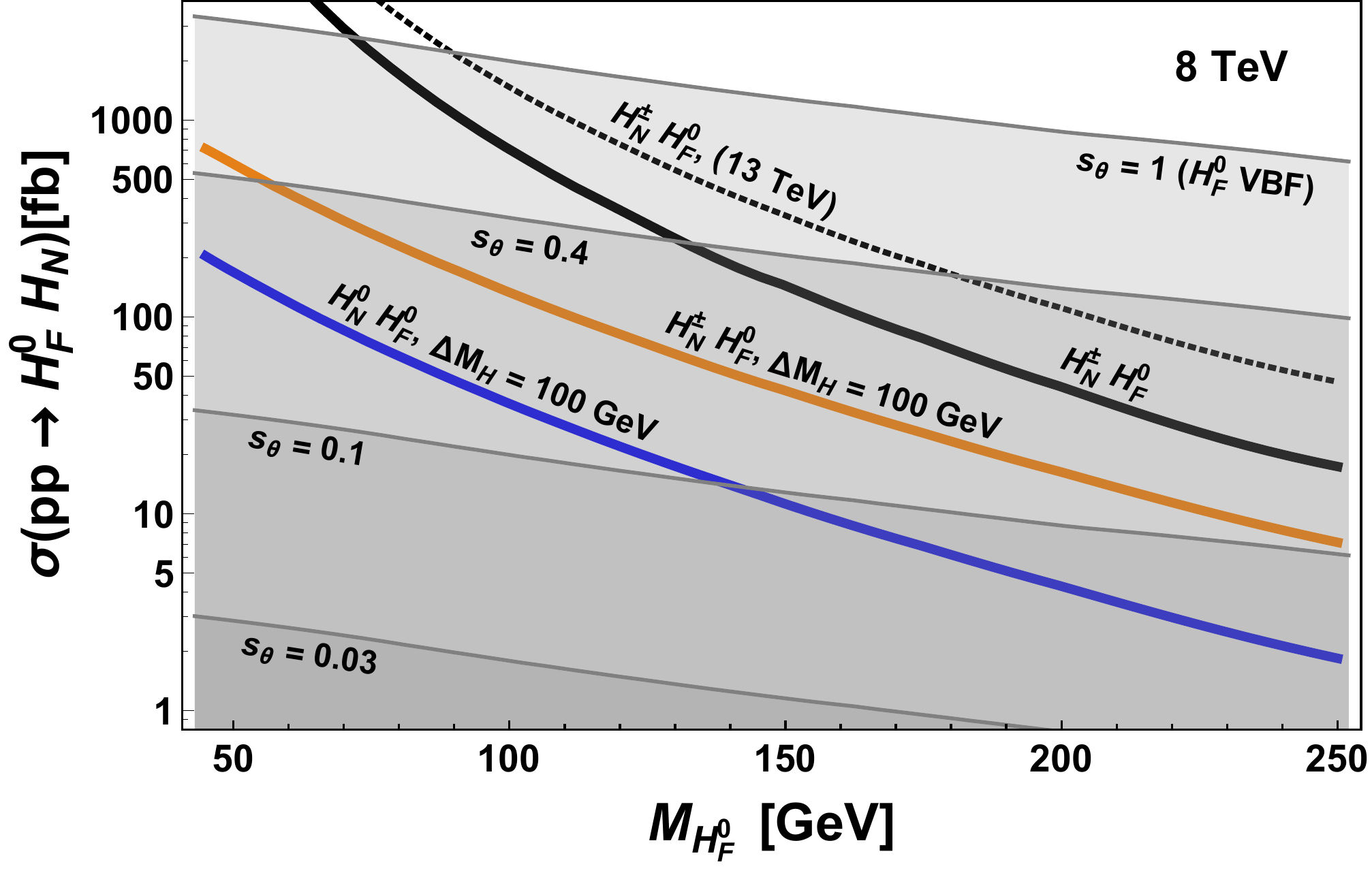}
\end{center}
\caption{Various Drell-Yan Higgs pair production cross sections for a neutral fermiophobic Higgs boson ($H_F^0$) at the LHC with $\sqrt{s} = 8$ TeV (thick solid curves) and $\sqrt{s} = 13$ TeV (black dotted curve).~The black curves show the production cross section for $H_F^0 H^{\pm}_N$ assuming degenerate masses.~Cross sections for DY Higgs pair production mediated by a $W$ (orange solid) or $Z$ (blue solid) when there is 100~GeV mass splitting ($M_{H^{\pm}_N} > M_{H^{0}_F}$) are also shown.~For comparison we show in the shaded gray region contours of the `vev mixing angle' $s_\theta$ as defined in~\eref{sth} for the VBF production channel where we have rescaled the 8 TeV SM cross sections~\cite{Dittmaier:2011ti,Dittmaier:2012vm,Heinemeyer:2013tqa} by $s_\theta^2$.} 
\label{fig:Hprod}
\end{figure}

The main focus of this study will be the $pp \to W^{\pm} \to H_F^0 H^{\pm}_N$ production channel for which we show the cross section (solid black) as a function of $H_F^0 $ mass assuming degenerate masses.~We see that it can be $\gtrsim \mathcal{O}(100)$~fb all the way up to $\sim 200$~GeV at 8 TeV while at 13 TeV (black dotted) it will be increased by roughly a factor of $\sim 2$.~If there is a $100$~GeV splitting between the neutral and charged scalars (solid orange) and assuming $M_{H^{\pm}_N} > M_{H^{0}_F}$ the cross section is considerably reduced, but still $\gtrsim \mathcal{O}(100)$~fb all the way up to $\sim 150$~GeV.~We also show for comparison the $Z$ mediated $H_F^0 H^{0}_N$ channel (blue solid curve) for the same mass splitting which we see is significantly smaller than the $W$ mediated channels, but again may be relevant for light masses.~Note there are also NLO contributions which may generate $\gtrsim \mathcal O(1)$ K-factors for Higgs pair production~\cite{Eichten:1984eu,Dawson:1998py,Degrande:2015xnm}, but we do not explore this issue here as it does not qualitatively affect our discussion. 

We also show for comparison in the gray shaded region the VBF cross section for $H_F^0$, which depends on the exotic Higgs vev.~We can parametrize this dependence generically through a SM doublet-exotic Higgs `vev mixing angles' ($c_\theta \equiv \cos\theta,~s_\theta \equiv \sin\theta$),
\bea\label{eq:sth}
c_\theta = \frac{v_h}{v},~
s_\theta = \frac{v_{ex}}{v}~~~(v = 246~\rm{GeV}),
\eea
where $v_h$ is the vev of the mostly SM Higgs doublet observed at 125~GeV and $v_{ex}$ represents schematically the sum (in quadrature), which may also include group theory factors, over all exotic Higgs vev contributions to EWSB.~So $s_\theta$ essentially parametrizes the relative contribution to the electroweak scale from the exotic Higgs sector.

With the definition in~\eref{sth} we can then obtain the VBF cross section by simply rescaling the 8 TeV SM prediction~\cite{Dittmaier:2011ti,Dittmaier:2012vm,Heinemeyer:2013tqa} by $s_\theta^2$ for which we show various contours.~These curves implicitly assume that the ratios of the $H_F^0$ couplings to $WW$ and $ZZ$ pairs equal those of the SM Higgs.~This will not be true for all Higgs bosons found in exotic Higgs sectors such as for example the custodial fiveplet in custodial Higgs triplet models~\cite{Georgi:1985nv,Hartling:2014zca,Cort:2013foa} to be examined in more detail below.~We see clearly that once the measurements of the Higgs boson at 125~GeV constrain $s_\theta \ll 1$, the VBF production channel quickly becomes highly suppressed relative to the DY Higgs pair production channels.~Similar behavior can be seen for the $VH$ production channels which are typically smaller than the VBF cross sections except at very low masses~\cite{Dittmaier:2011ti,Dittmaier:2012vm,Heinemeyer:2013tqa}.

To summarize, we see that $\gtrsim\mathcal{O}(100)$~fb cross sections are obtained for the $pp \to H_F^0 H_N^{\pm}$ Higgs pair production channel in the mass range $45 - 250$~GeV.~Crucially this production mechanism is present even in the limit of vanishing exotic Higgs vev unlike VBF and VH production.~As we will see, diphoton and diboson searches at the 8 TeV are sensitive to  $\lesssim\mathcal{O}(100)$~fb cross section times branching ratios.~Thus if the branching ratios to dibosons are large, searches at the LHC for pairs of photons or $Z$ and $W$ bosons should be able to probe fermiophobic Higgs bosons in this mass range.

\subsection{Fermiophobic Higgs Diboson Decays}

In addition to the $WHH$ vertex in~\eref{gwhh}, $H_F^0$ will have couplings to $WW$ and $ZZ$ pairs which are generated during EWSB and which will be proportional to the \emph{exotic} Higgs vev~\cite{Georgi:1985nv,Akeroyd:2003bt,Akeroyd:2010eg,Cort:2013foa,Hartling:2014zca}.~We can parametrize these couplings generically with the following lagrangian,
\bea
\label{eq:LZW}
\mathcal{L}
&\supset&
s_\theta
\frac{H_F^0}{v} 
\Big( g_{Z} m_Z^2 Z^\mu Z_\mu + 2 g_{W} m_W^2 W^{\mu+} W^-_{\mu} 
\Big) ,
\eea
where $g_Z$ and $g_W$ are fixed by the $SU(2)_L \otimes U(1)_Y$ representation to which $H_F^0$ belongs.~The factor of $s_\theta$ defined in~\eref{sth} ensures that as the exotic Higgs vev tends to zero (i.e. $s_\theta \to 0$) the $H_F^0 VV$ couplings vanish along with the VBF and VH production mechanisms.~Again we assume we are in an `alignment' limit so that no Higgs mass mixing angles enter into~\eref{LZW}.~However, even in the case where they do, this dependence largely cancels when considering branching ratios since it enters as an overall factor along with the `vev mixing angle' $s_\theta$.~The ratio of the $g_Z$ and $g_W$ couplings,
\bea\label{eq:lamWZ}
\lambda_{WZ} = g_{W}/g_{Z},
\eea
is an important quantity and is fixed by custodial symmetry at tree level to be $|\lambda_{WZ}| = 1$ or $|\lambda_{WZ}| = 1/2$ for a custodial singlet and fiveplet respectively~\cite{Low:2010jp}.~Though sizeable deviations from these two values are in principle possible, they are difficult to reconcile with electroweak precision data in a natural way.~Therefore, in what follows we will consider only these two cases.~Note also that a factor of $s_\theta$ has been implicitly canceled in~\eref{lamWZ}.

At one loop the $g_W$ couplings in~\eref{LZW} will also generate effective couplings to $\gamma\gamma$ and $Z\gamma$ pairs via the $W$ boson loops shown in~\fref{HtoVA}.~We can parametrize these couplings with the effective operators,
\bea
\label{eq:LZA}
\mathcal{L}
&\supset&
\frac{H_F^0}{v} 
\Big( 
\frac{c_{\gamma\gamma} }{4} 
F^{\mu\nu} F_{\mu\nu} +
\frac{c_{Z\gamma}}{2} 
Z^{\mu\nu} F_{\mu\nu} 
\Big) ,
\eea
where $V_{\mu\nu}=\partial_\mu V_\nu - \partial_\nu V_\mu$.~We again define similar ratios,
\bea\label{eq:lamVA}
\lambda_{V\gamma} = c_{V\gamma}/g_{Z},
\eea
where $V = Z, \gamma$ and we have implicitly absorbed a factor of $s_\theta$ into $g_Z$.~There are also contributions to the effective couplings in~\eref{LZA} from the additional charged Higgs bosons which are necessarily present.~Depending on the Higgs potential, there may be dimensional parameters entering in the trilinear scalar couplings~\cite{Georgi:1985nv,Akeroyd:2003bt,Akeroyd:2010eg,Cort:2013foa,Hartling:2014zca} which contribute to the charged scalar loop amplitude and which are independent of the exotic vev ($s_\theta$).~In these cases one can easily obtain larger values of $\lambda_{V\gamma}$ either by taking this new mass scale large compared to the weak scale or by taking the limit $s_\theta \ll 1$, thus suppressing the tree level coupling to $ZZ$ and $WW$.~In this case loop induced decays to $WW$ and $ZZ$ can also become relevant.~The effective couplings in~\eref{LZA} can also be enhanced when the loop particles carry large charges and interfere constructivly with the $W$ boson loop contribution~\cite{Akeroyd:2012ms}.~However, these contributions could in principle conspire to cancel~\cite{Arhrib:2011vc,Kanemura:2012rs} leading to small $c_{V\gamma}$ effective couplings.
%
\begin{figure}[tbh]
\begin{center}
\includegraphics[scale=.6]{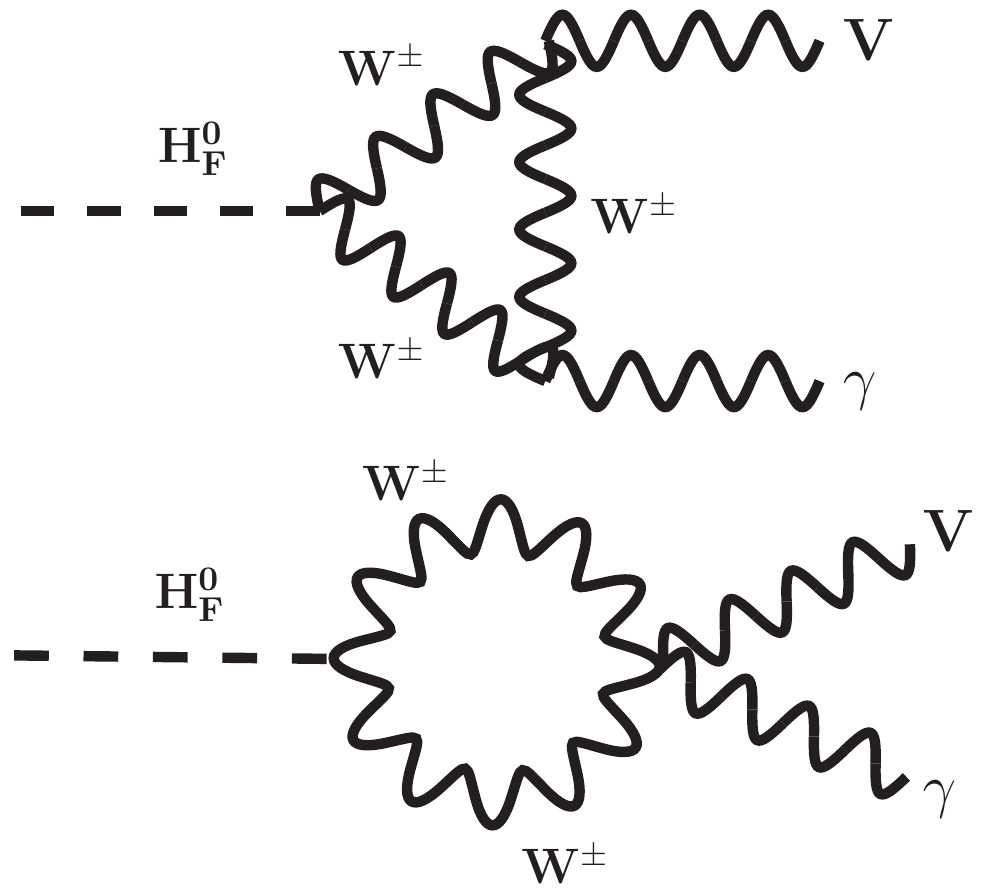}
\end{center}
\caption{One loop contributions from $W$ boson loops to the $H_F^0 V\gamma$ ($V = Z, \gamma$) effective couplings defined in~\eref{LZA}.}
\label{fig:HtoVA}
\end{figure}

If there are no exotic states light enough to decay into, and since there is no $b\bar{b}$ decay to compete with, the neutral fermiophobic Higgs bosons will decay almost entirely into electroweak gauge boson pairs and in particular photons at low masses~\cite{Mrenna:2000qh,Landsberg:2000ht,Arhrib:2003ph,Akeroyd:2007yh,Akeroyd:2012ms}.~Loop mediated decays to light SM fermions can occur thus violating the fermiophobic condition, but will be suppressed by the fermion masses and furthermore must be fixed by renormalization~\cite{Diaz:1994pk,Akeroyd:1995hg} in certain cases.~Here we will assume the fermiophobic condition is maintained by either an appropriately chosen renormalization condition~\cite{Barroso:1999bf,Brucher:1999tx} or via a symmetry~\cite{Akeroyd:2010eg} such as custodial symmetry~\cite{Georgi:1985nv,Hartling:2014zca,Cort:2013foa}.~Under these assumptions the branching ratios of $H_F^0$ will only depend on the ratios in~\eref{lamWZ} and~\eref{lamVA}, and in some cases only on $\lambda_{WZ}$ if the $W$ loop (see~\fref{HtoVA}) dominates the $H_F^0V\gamma$ effective couplings.~In this case any $s_\theta$ dependence in $\lambda_{V\gamma}$ cancels explicitly.

Since the qualitative behavior of the branching ratios is largely dominated by phase space considerations, they will share many features in any fermiophobic Higgs model.~At low masses, below $\sim 120 - 150$~GeV, the branching ratio into pairs of photons starts to become significant and quickly dominant below the $W$ mass, or at higher masses if the couplings to photons are enhanced.~At larger masses the three and four body decays involving $W$ and $Z$ bosons become relevant and eventually completely dominant above the $WW$ and $ZZ$ thresholds.~At even higher masses, either the $ZZ$ or $WW$ branching ratio can be the largest decay mode depending on the value of the ratio of the couplings, $\lambda_{WZ}$.

We illustrate these features in~\fref{HBR} where we show branching ratios for two different fermiophobic Higgs scenarios in the mass range $45 - 150$~GeV.~In both cases we take $|\lambda_{WZ}| = 1$, which is possible in all two Higgs doublet as well as Higgs triplet models.~To obtain the three and four body decays we have integrated the analytic expressions for the $H_F^0 \to V\gamma \to 2\ell\gamma$ and $H_F^0 \to VV \to 4\ell$ fully differential decay widths computed and validated in~\cite{Chen:2012jy,Chen:2013ejz,Chen:2014ona}.~For the explicit $W$ loop functions which contribute to the effective couplings we use the parametrization and implementation found in~\cite{Chen:2015rha}.
\begin{figure}[tbh]
\begin{center}
\includegraphics[scale=.48]{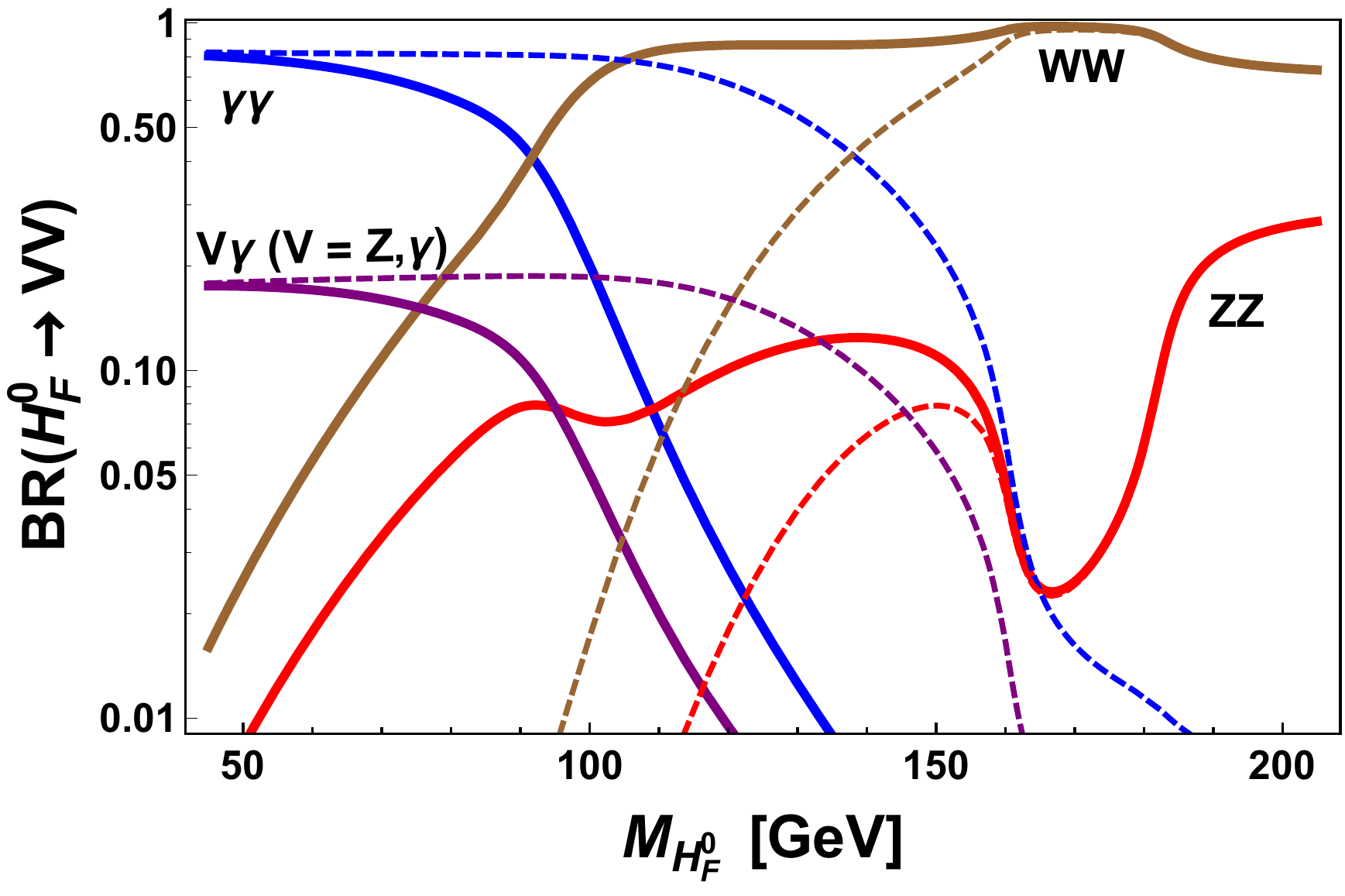}
\end{center}
\caption{Branching ratios for $H_F^0$ as a function of its mass where for all curves we have set $|\lambda_{WZ}| = 1$ (see~\eref{lamWZ}).~For the solid curves we have assumed the couplings to $\gamma\gamma$ and $Z\gamma$ are generated only by the $W$ boson loop in~\fref{HtoVA}.~For the dashed curves we have taken the effective couplings to $\gamma\gamma$ and $Z\gamma$ as free paramaters and set $\lambda_{\gamma\gamma} = \lambda_{Z\gamma} = 0.05$ (see~\eref{lamVA}).}
\label{fig:HBR}
\end{figure}

The first scenario (solid curves) assumes the fermiophobic Higgs interactions are dominated by the couplings in~\eref{LZW}.~In this case the effective couplings to $Z\gamma$ and $\gamma\gamma$ are generated only by the $W$ loop shown in~\fref{HtoVA}.~This case has been considered previously in~\cite{Stange:1993ya,Diaz:1994pk}, but did not explicitly include the virtual photon contribution in $H_F^0 \to V^\ast\gamma \to 2f\gamma$ (purple curves) which is dominated by the $\gamma^\ast\gamma$ component at low masses and can be as large as $\mathcal{O}(20\%)$.~While the size of this contribution depends on experimental phase space cuts, as emphasized in~\cite{Chen:2012jy,Chen:2013ejz,Chen:2014gka,Chen:2014ona,Chen:2015iha,Chen:2015rha,Stolarski:2016dpa}, virtual diphoton effects can provide valuable information in scalar decays.~Note there is also the two body $H_F^0 \to Z\gamma$ decay, but it is less than $1\%$ over this mass range.~We also emphasize that in this case all of the $H \to VV$ decay amplitudes depend linearly on the exotic Higgs vev (or $s_\theta$) and thus the branching fractions will be independent of the vev.~As we will discuss below, in some cases this vev independence of the branching ratios can be utilized, along with the Higgs pair production mechanism, to obtain constraints on fermiophobic scalars which are independent of the vev.

In the second scenario (dashed curves) we consider the possibility of generating large effective coupling to $\gamma\gamma$ and $Z\gamma$ by taking $\lambda_{V\gamma} = \lambda_{\gamma\gamma} = \lambda_{Z\gamma} = 0.05$.~This is to be compared to $\lambda_{V\gamma} \sim 0.005 - 0.01$ from only the $W$ loop contribution which depends on the mass of $H_F^0$.~As discussed above, such large values for this ratio can easily be obtained\footnote{As an explicit example if we take $H_F^0$ to be $\sim 160$~GeV and to be the neutral component of the custodial fiveplet scalar found in custodial Higgs triplet models~\cite{Georgi:1985nv,Akeroyd:2003bt,Akeroyd:2010eg,Cort:2013foa,Hartling:2014zca}, we find that for trilinear couplings $A \sim 15\, s_\theta$~TeV, ratios of $\lambda_{\gamma\gamma} \sim 0.05$ can be easily obtained via the contribution from its (degenerate) doubly charged component.~Note that such light masses are not ruled out by previous searches for the doubly (or singly) charged component when $s_\theta \lesssim 0.3$~\cite{Logan:2015xpa}.}~in the limit $s_\theta \ll 1$ if there exist additional mass scales apart from the Higgs vevs in the scalar potential or if the loop particles carry large charges.~We see that in this case of enhanced couplings to photons the diphoton channel can be sizable all the way up to the $WW$ threshold.~We also see the $H_F^0 \to V\gamma \to 2f\gamma$ three body decay through an off-shell photon or $Z$ can also be sizable for masses up to $\sim 130$~GeV and may be interesting to study further.

Depending on how these large effective couplings are generated, there may be a dependence on the exotic Higgs vev introduced into the branching ratios.~However, even in this case the branching ratios are still largely independent of the vev since over much of the mass range either the $\gamma\gamma$ (and $\gamma^\ast\gamma$) decay dominates, or $WW$ and $ZZ$ decays dominate.~The same holds true if the effective couplings to $\gamma\gamma$ and $Z\gamma$ are highly suppressed due to cancellations.~Thus one can again obtain limits on fermiophobic Higgs bosons which are independent of their vevs.~However, it would be interesting to consider a detailed analysis of $H_F^0$ masses above the $Z$ mass and below the $WW$ threshold where all decays can in principle be sizeable simultaneously and where the vev dependence can be non-negligible.

In both cases considered in~\fref{HBR} we see the universal features of a fermiophobic Higgs boson.~Namely, large branching ratios into photons at lower masses and large branching ratios to $WW$ and $ZZ$ at larger masses.~As we will demonstrate below, these diboson decays can be combined along with the Higgs pair production mechanism to provide stringent constraints on a variety of fermiophobic Higgs scenarios and, in some cases, these constraints being independent of the exotic Higgs vev or Higgs mixing effects.

\section{Probing Fermiophobic Higgs\\ 
~~~~~~Bosons at the LHC}\label{sec:LHC}

With this discussion of fermiophobic Higgs boson production and decays in mind, we now examine the possibility of using diphoton and diboson searches at the LHC to search for fermiophobic Higgs bosons.~In what follows we will consider only the $pp \to W^{\pm} \to H_N^{\pm} H_F^0$ Higgs pair production mechanism shown in~\fref{HHprod}, where $H_F^0$ and $H_N^{\pm}$ can be degenerate or have a large mass splitting.~We again consider the `normalized' $H_F^0 H^{\pm}_N$ Higgs pair production channels discussed in~\fref{Hprod} where any potential Higgs mixing angle and group theory factor $C_N$ has been factored out of the $WHH$ vertex in~\eref{gwhh}.~This can also be considered as an assumption of `alignment' between the gauge and mass eigenstates and setting $C_N = 1$, or the large $\tan\beta$ plus fermiophobic limit~\cite{Akeroyd:2003bt} in two Higgs doublet models.~We assume the fermiophobic condition is maintained at loop level by either an appropriate renormalization condition~\cite{Barroso:1999bf,Brucher:1999tx} or global symmetry~\cite{Georgi:1985nv,Akeroyd:2010eg,Cort:2013foa,Hartling:2014zca}.~Furthermore, we assume that $H_F^0$ cannot decay to any exotic states and, for all results shown in this section, we take $|\lambda_{WZ}| = 1$.

We emphasize that this $pp \to W^{\pm} \to H_N^{\pm} H_F^0$ pair production mechanism had not been considered in Higgs searches until a very recent CDF analysis~\cite{Aaltonen:2016fnw} of Tevatron data in the $4\gamma + X$ channel.~Here we examine its utility to search for fermiophobic Higgs bosons at the 8 TeV LHC in the diphoton~\footnote{We have also considered diphoton searches at Tevatron via $p\bar{p} \to H_F^0 H^{\pm}_N$ production, but find that the Higgs pair production cross sections are too small to utilize diphoton limits to search for fermiophobic Higgs bosons.}, $ZZ$, and $WW$ channels.~These are the first searches to constrain a fermiophobic Higgs boson with a small vev when VBF and VH production become highly suppressed.

\subsection{Diphoton Probes for Light Masses}\label{eq:diphoton}

We first consider the possibility of probing fermiophobic Higgs bosons with diphoton searches at 8 TeV LHC.~We show in~\fref{AAprodxbr} the production cross section times branching ratio into diphotons (blue curves) as a function of $H_F^0 $ mass for a variety of fermiophobic Higgs scenarios.~We show the $95\%$ exclusion limit (black dashed) from~8 TeV ATLAS diphoton searches~\cite{Aad:2014ioa} for masses above $65$~GeV.~In order to illustrate the loss of sensitivity in VBF based searches for small exotic Higgs vevs, we also show (gray shaded region) contours of $s_\theta$ (see~\eref{sth}) for single $H_F^0$ VBF production at 8 TeV LHC obtained by rescaling the SM cross sections~\cite{Dittmaier:2011ti,Dittmaier:2012vm,Heinemeyer:2013tqa}.
\begin{figure}[tbh]
\begin{center}
\includegraphics[scale=.47]{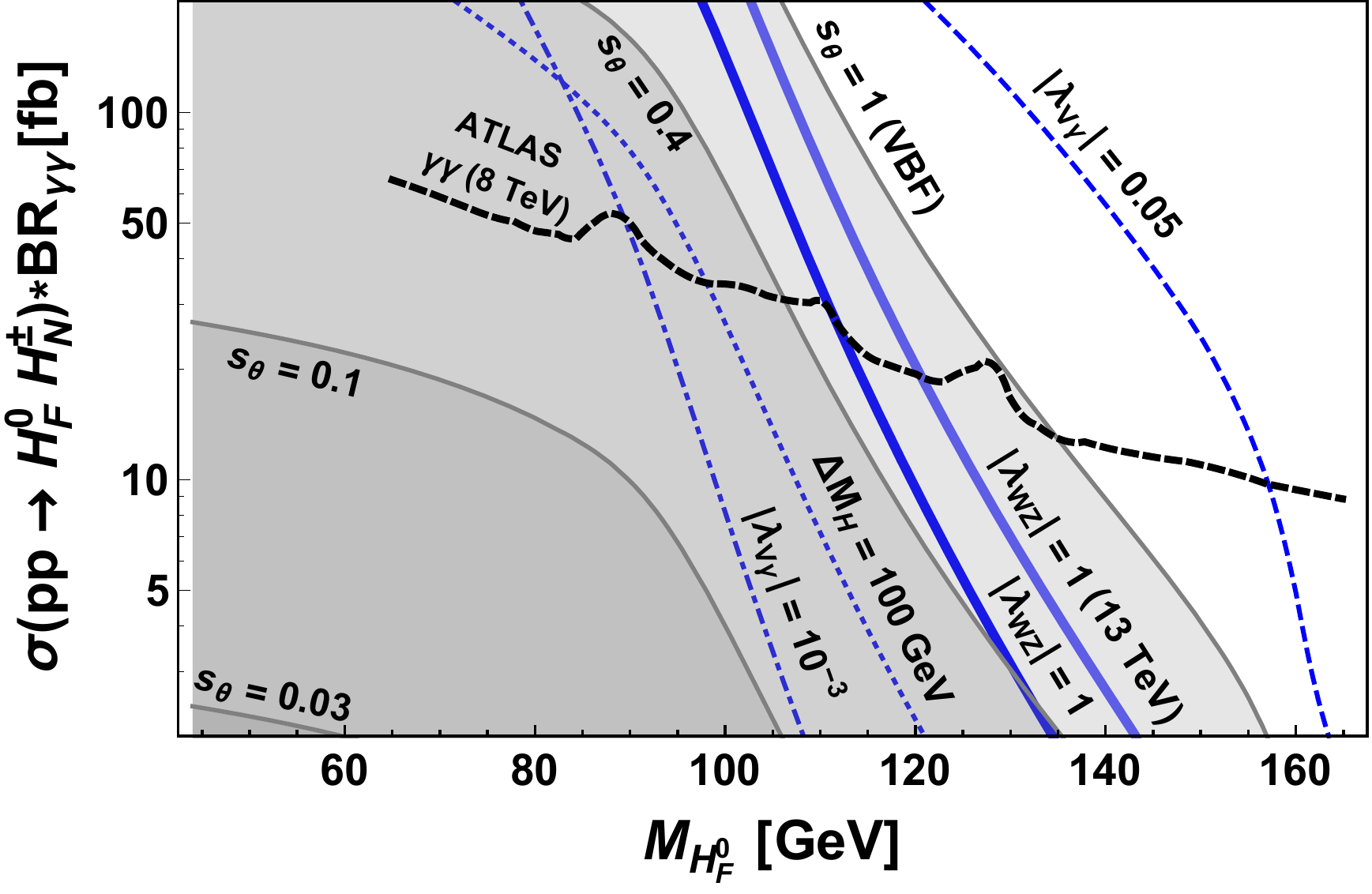}
\end{center}
\caption{Production cross sections times branching ratio into diphotons (blue curves) at 8 TeV for a variety of fermiophobic Higgs scenarios assuming the $H_F^0 H^{\pm}_N$ Higgs pair production channels discussed in~\fref{Hprod}.~For all curves we assume $|\lambda_{WZ}| = 1$ (see~\eref{lamWZ}).~The $95\%$ exclusion limit (black dashed) from~8 TeV ATLAS diphoton searches~\cite{Aad:2014ioa} is also shown.~In the gray shaded region we show contours of $s_\theta$ (see~\eref{sth}) for single $H_F^0$ VBF production.~See text for more information.
}
\label{fig:AAprodxbr}
\end{figure}
%

We first note that for the case of $H_F^0 H^{\pm}_N$ production, assuming degenerate masses and dominant $W$ boson loop (solid thick blue curve), fermiophobic Higgs masses below around $\sim 115$~GeV can be ruled out.~Again for comparison we show the same production channel at 13 TeV (light blue solid curve) where we see masses up to $\sim 125$~GeV could be probed and perhaps ruled out once 13 TeV diphoton data becomes available.~For a specific model this bound may be higher or lower depending on the group theory factor $C_N$ entering the coefficient in~\eref{gwhh} and assuming small Higgs mass mixing effects, but this generally will not drastically change the bound as we will see below for custodial Higgs triplet models.~We emphasize that the recent CDF analysis~\cite{Aaltonen:2016fnw} of multiphoton Tevatron data is insensitive to this degenerate case.~For the case of a $100$~GeV splitting (blue dotted curve) between the neutral and charged scalars ($M_{H^{\pm}} > M_{H^{0}}$) the limit at 8 TeV is reduced to $\sim 100$~GeV~\footnote{Note the parameter point $M_{H_F^0} = 100, M_{H^{\pm}} = 200$~GeV is not ruled out by Tevatron data~\cite{Aaltonen:2016fnw}.}.~While we have not shown a case where $M_{H^{\pm}} < M_{H^{0}}$, clearly diphoton searches can also be applied in this case,~and in particular when the mass splitting is much less than the $W$ mass leading to a suppression of the $H_F^0 \to H^{\pm} W^{\mp}$ decay.

Again we consider the possibility of constructive interference effects generating enhanced (by an order of magnitude) effective coupling to $\gamma\gamma$ and $Z\gamma$ by taking $\lambda_{V\gamma} = 0.05$ (blue dashed curve).~We see that in this case the diphoton channel can potentially rule out fermiophobic Higgs boson masses all the way up to around the $WW$ threshold.~We also consider the case where interference effects conspire to cancel giving small effective $V\gamma$ couplings by taking $ \lambda_{V\gamma} = 10^{-3}$ (blue dot dashed curve).~In this case the limits are noticeably reduced, but nevertheless masses below $\sim 90$~GeV can still be ruled out.~We also see for the VBF production mode that if values of $s_\theta \approx 1$ were still allowed by measurements of the 125~GeV Higgs boson, they would be ruled out by this diphoton search for masses below $\sim 140$~GeV.~Once $s_\theta$ is constrained to be $\lesssim 0.4$, the VBF production mode becomes less sensitive than the Higgs pair production mode and totally irrelevant for $s_\theta \lesssim 0.1$.

We also emphasize the importance of extending diphoton searches below 65~GeV since masses in this range are not ruled out by Tevatron searches for small mass splittings between the Higgs pair.~There are in principle limits on the charged Higgs boson from LEP which apply, but these can be evaded if the charge Higgs is also fermiophobic.~Depending on how the charged Higgs decays there may still be relevant constraints, but a detailed investigation is beyond the scope of this work.

Generally speaking there are a few ways to evade the limits discussed here.~The first is to introduce large amounts of Higgs mass mixing via multiple neutral CP even scalars.~Since mass mixing with the SM-like Higgs at 125~GeV is constrained to be small~\cite{Khachatryan:2014kca}, the mixing must be among \emph{exotic} scalars to give large effects and qualitatively affect our results.~In some models, this mass mixing is forbidden by global symmetries as for example in custodial Higgs triplet models to be discussed more below.~An additional way to evade these limits is to engineer for even more precise cancellations among the contributions to the $H_F^0\gamma\gamma$ effective coupling leading to even more suppressed values than $10^{-3}$ for $\lambda_{V\gamma}$.~Finally, if there is no additional charged or neutral Higgs light enough to be appreciably pair produced along with $H_F^0$ then production cross sections become highly suppressed making $H_F^0$ difficult to observe.~Possibilities to obtain very robust bounds and test extreme regions of parameter space in fermiophobic Higgs models at a 100~TeV collider would also be interesting to consider.

\subsection{Diboson Probes of Intermediate Masses}
We now consider the possibility of combining the $p\bar{p} \to H_F^0 H^{\pm}_N$ production mechanism with $WW$ and $ZZ$ decays in order to study fermiophobic Higgs bosons at the LHC.~We show in~\fref{VVprodxbr} the production cross section times branching ratio into $WW$ (top) and $ZZ$ (bottom) as a function of the $H_F^0 $ mass for the various fermiophobic Higgs scenarios discussed in~\fref{AAprodxbr}.~The $95\%$ exclusion limits (black dashed) from $7 + 8$~TeV $WW$ and $ZZ$ searches at CMS~\cite{Khachatryan:2015cwa} are also shown for each channel.~We see that limits from $7 + 8$ TeV $WW$ and $ZZ$ searches are not quite sensitive to these fermiophobic Higgs scenarios.~However, if 13 TeV data can improve upon current limits by roughly an order of magnitude these channels should become promising probes of fermiophobic Higgs bosons with masses above where diphoton searches are sensitive and potentially up to $\sim 250$~GeV.
%
\begin{figure}[tbh]
\begin{center}
\includegraphics[scale=.485]{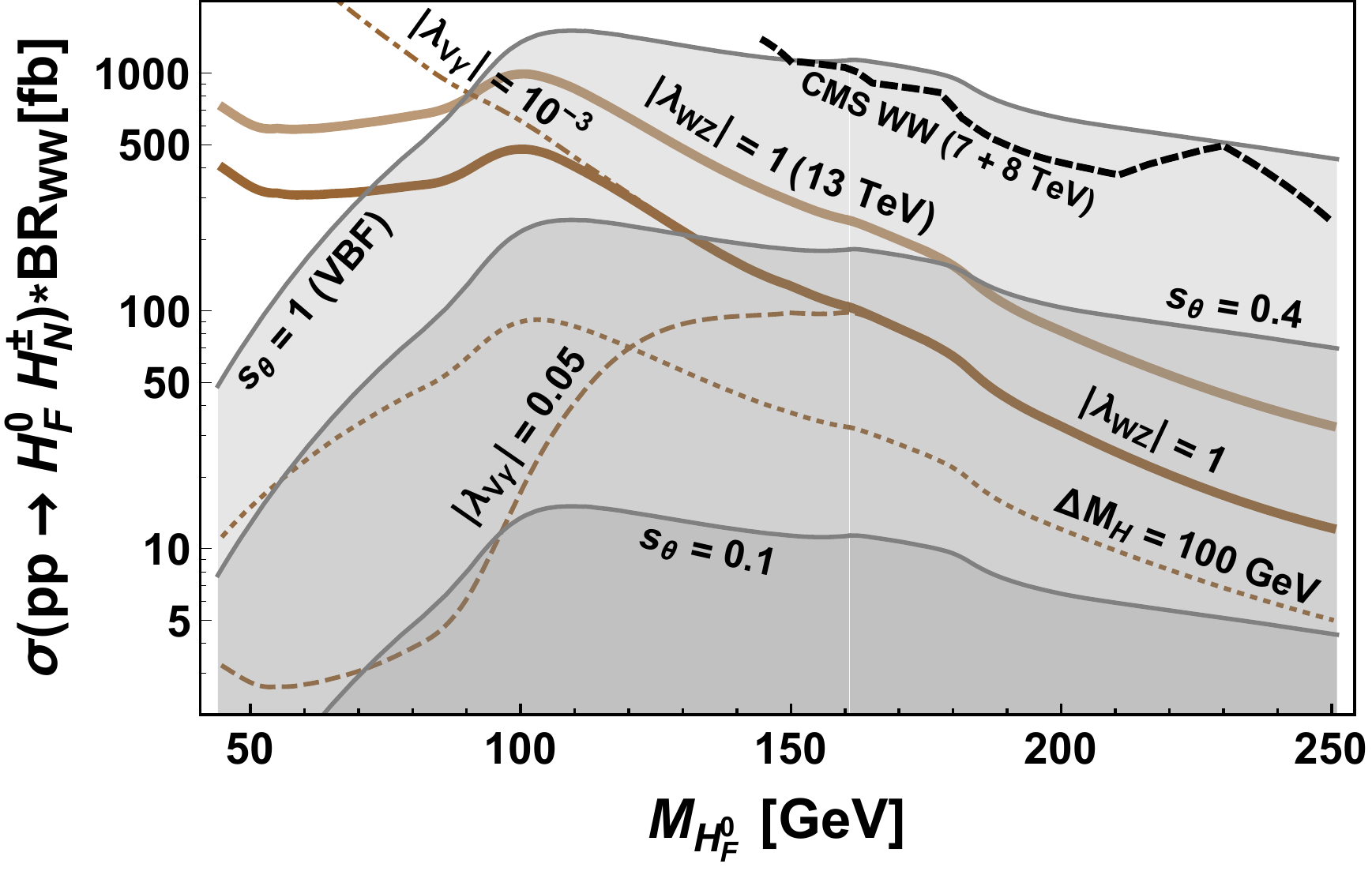}
\includegraphics[scale=.47]{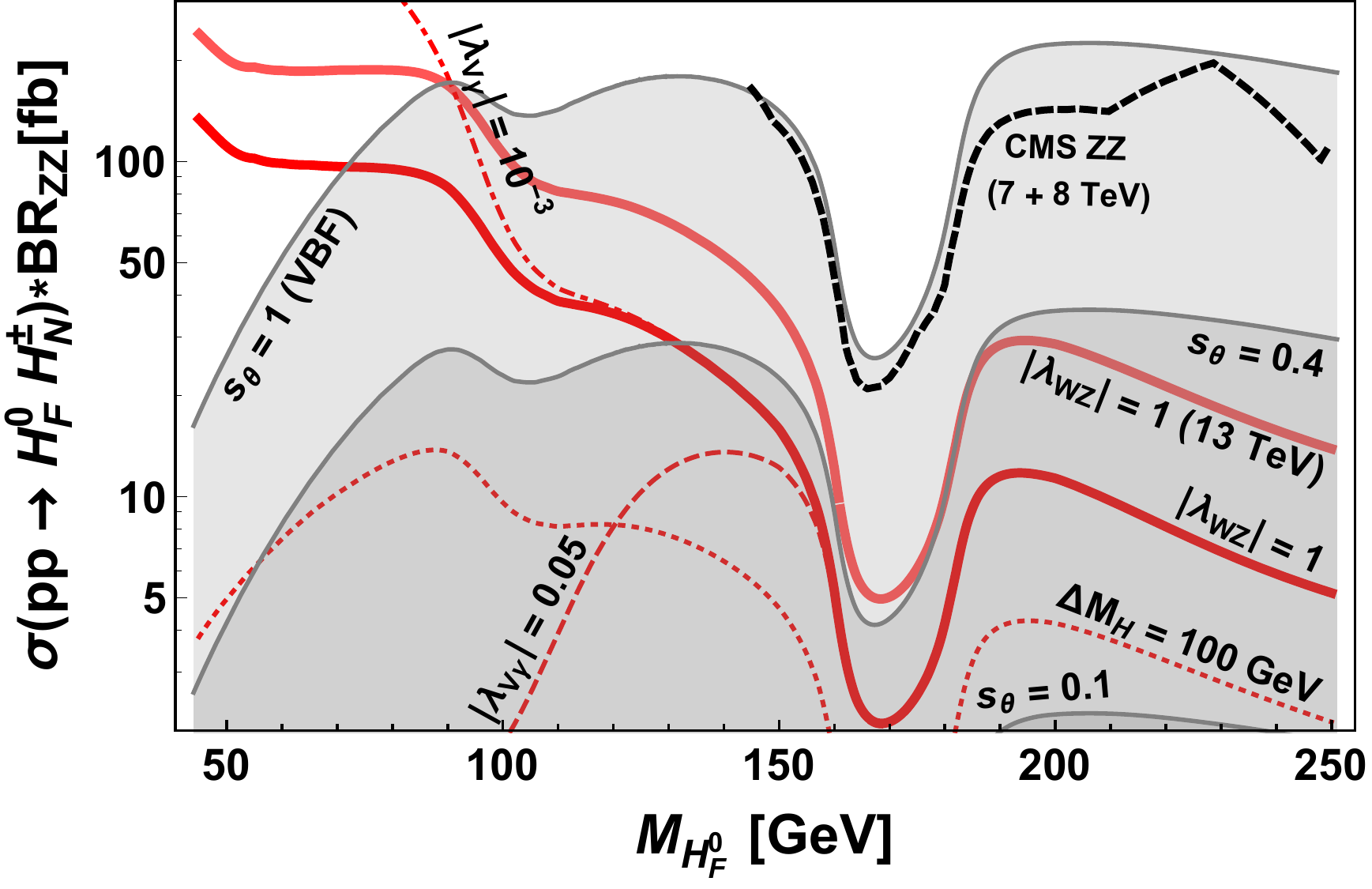}
\end{center}
\caption{Same as~\fref{AAprodxbr}, but for WW (top) and ZZ (bottom) decay channels.~The $95\%$ exclusion limits (black dashed) from $7 + 8$~TeV $WW$ and $ZZ$ searches at CMS~\cite{Khachatryan:2015cwa} are also shown.}
\label{fig:VVprodxbr}
\end{figure}
%

Furthermore, we emphasize that these $WW$ and $ZZ$ searches should be extended to lower masses below the $VV$ threshold, where cross sections can be larger and where there are currently no relevant $WW$ or $ZZ$ searches for a fermiophobic Higgs with a small vev.~In particular, by extending these searches to lower masses, we have an additional probe which may allow us to uncover light fermiophobic Higgs bosons with suppressed couplings to photons (dot-dashed curves) where diphoton searches lose sensitivity.~On the other hand if the couplings to photons are enhanced these $WW$ and $ZZ$ searches become less sensitive.~We also note that a dedicated analysis of masses around the 125~GeV Higgs could be particularly interesting as this region has been neglected in terms of exotic Higgs searches since the discovery of the SM-like Higgs boson.~Furthermore, in this region all decay channels are potentially sensitive.

\section{Closing the `Fiveplet Window'}\label{sec:H5}

As an explicit example of a fermiophobic scalar
sector which contributes to EWSB, we consider custodial Higgs triplet
models, which consist of the SM (or MSSM) Higgs sector plus three custodial electroweak triplet scalars.~There are a number of variations of custodial Higgs triplet models, both non-supersymmetric~\cite{Georgi:1985nv,Chanowitz:1985ug,Gunion:1989ci,Gunion:1990dt,Gunion:1989ci,Gunion:1990dt,Aoki:2007ah,Chiang:2012cn,Chiang:2013rua,Hartling:2014zca,Logan:2015xpa} and supersymmetric~\cite{Cort:2013foa,Garcia-Pepin:2014yfa,Delgado:2015aha,Delgado:2015bwa}, but their differences are not relevant for our current study.~The
crucial feature that all of them share, in addition to being easily made to satisfy constraints from electroweak precision data, is that after EWSB the Higgs triplets decompose into representations of the custodial $SU(2)_C$ global symmetry.~In particular, all custodial Higgs triplet models contain a fermiophobic scalar ($H_5$) transforming as a fiveplet under the
custodial symmetry and which has a CP even neutral ($H_5^0$), singly ($H_5^\pm$), and doubly ($H_5^{\pm\pm}$) charged components with degenerate masses.

Custodial symmetry also prevents the neutral component from mixing with other neutral scalars and in particular with the 125~GeV SM Higgs boson, allowing for the fermiophobic condition to be maintained~\footnote{Due to hypercharge interactions, custodial breaking effects are introduced at one loop which can spoil the fermiophobic and degenerate mass conditions for the custodial fiveplet. But these effects are naturally small~\cite{Gunion:1990dt,Garcia-Pepin:2014yfa} allowing for these conditions to be maintained to a good approximation.} without fine tuning~\cite{Georgi:1985nv,Cort:2013foa,Hartling:2014zca}, in contrast to two Higgs doublet models~\cite{Akeroyd:2010eg}.~Thus no mixing angles enter in~\eref{gwhh} for the $W^{\mp}H_5^0 H_5^\pm$ vertex while the group theory factor is fixed in all custodial Higgs triplet models to be $C_N = \sqrt{3}/2$, as is the ratio of $WW$ and $ZZ$ couplings~\cite{Low:2010jp} at $|\lambda_{WZ}| = 1/2$ (see~\eref{lamWZ}).~We also emphasize that there is no dependence on the Higgs triplet vev~\cite{Hartling:2014zca} in the $W^{\mp}H_5^0 H_5^\pm$ vertex~\cite{Georgi:1985nv,Akeroyd:2010eg,Cort:2013foa,Hartling:2014zca}.~Combined with the largely vev independent branching ratios, this allows us for the first time to use diphoton and diboson searches at the LHC to put robust limits on custodial fiveplet scalars which are \emph{independent} of the Higgs triplet vev.

In~\fref{prodxbr} we show the $pp \to W^{\pm} \to H_5^{\pm} H_5^0$ production cross section times branching ratio at 8 TeV (solid curves) for a custodial fiveplet decay into photon (blue), $WW$ (brown), and $ZZ$ (red) pairs at 8 TeV LHC.~We also show the limits (dashed lines) coming from ATLAS diphoton searches at 8 TeV~\cite{Aad:2014ioa} (blue) as well as CMS $7 + 8$~TeV searches~\cite{Khachatryan:2015cwa} for decays to $WW$ (brown) and $ZZ$ (red).~Our leading order results for the $pp \to W^{\pm} \to H_5^{\pm} H_5^0$ production cross sections are calculated using the Madgraph/GM model implementation from~\cite{Alwall:2014hca,Hartling:2014xma}.~The branching ratios are obtained from the partial widths into $\gamma\gamma, V^\ast\gamma~(V = Z, \gamma), WW$, and $ZZ$ which are computed for the mass range $45 - 250$~GeV.~They have a similar behavior as those in~\fref{HBR} except that at high mass $ZZ$ dominates due to the fact that $\lambda_{WZ} = 1/2$~\cite{Chiang:2015kka}.~The relevant three and four body decays are obtained by integration of the analytic expressions for the $H_5^0 \to V\gamma \to 2\ell\gamma$ and $H_5^0 \to VV \to 4\ell$ fully differential decay widths computed in~\cite{Chen:2012jy,Chen:2013ejz,Chen:2014ona}.~We note that these branching ratios include the $\gamma^\ast\gamma$ contribution which, as shown in~\fref{HBR}, can be sizeable at low masses.

We focus on the regime where the effective couplings of the fiveplet to $\gamma\gamma$ and $Z\gamma$ are dominated by the $W$ loop contribution shown in~\fref{HtoVA}.~The effects of the charged scalar sector could in principle be large~\cite{Akeroyd:2012ms} leading to enhanced or suppressed effective couplings to photons.~As discussed above, and shown in~\fref{AAprodxbr}, this can affect the upper limit of masses which can be ruled out and could in principle allow for masses up to the $WW$ threshold to be ruled out by diphoton searches.~Since these effects are more model dependent we do not consider them here.
%
\begin{figure}[tbh]
\begin{center}
\includegraphics[scale=.51]{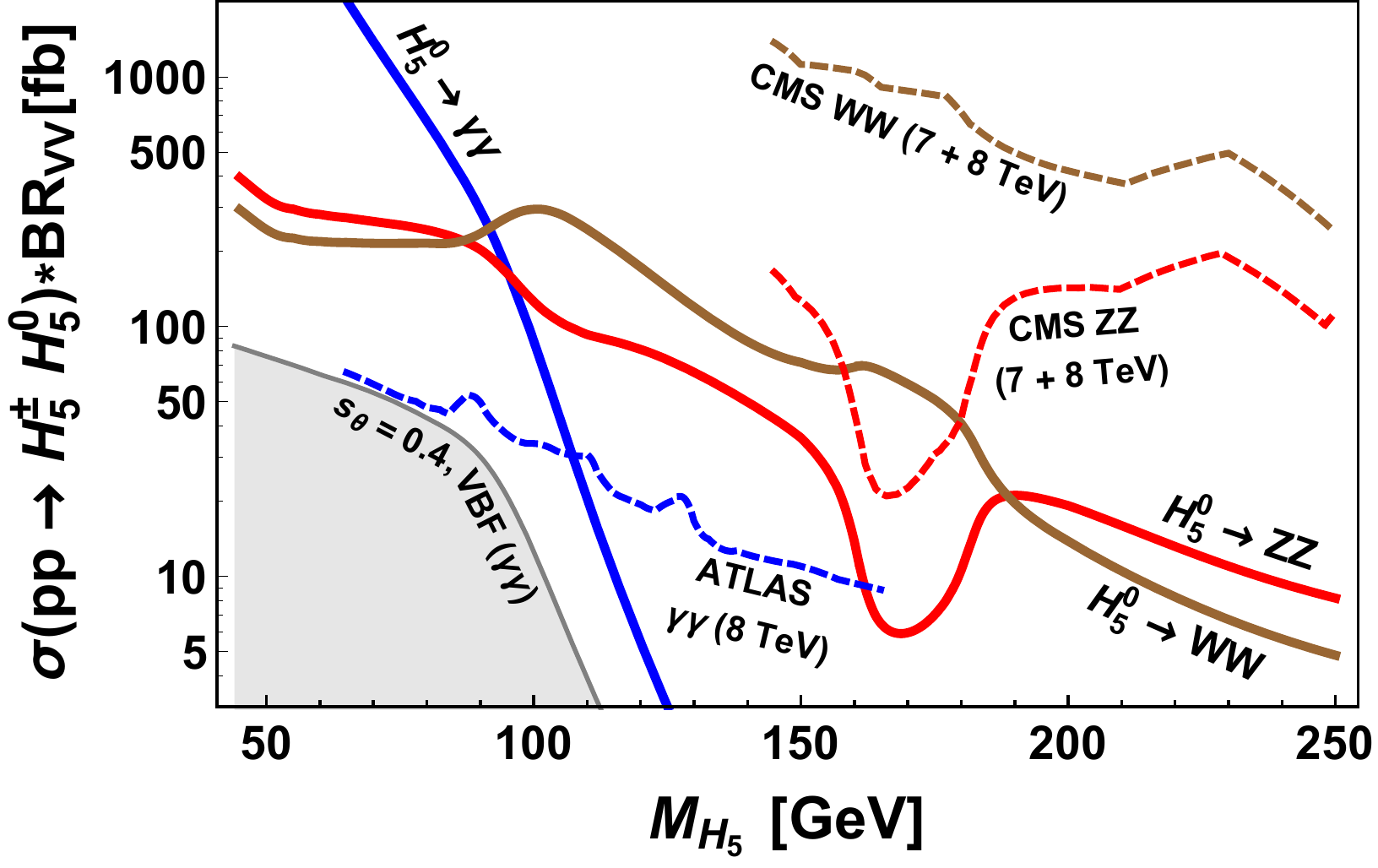}
\end{center}
\caption{Drell-Yan $H^{0}_5 H^{\pm}_5$ production cross sections times branching ratio at 8 TeV (solid curves) into $\gamma\gamma$ (blue), $ZZ$ (red), and $WW$ (brown) for the fermiophobic fiveplet found in custodial Higgs triplet models.~The $95\%$ exclusion limits (dashed curves) from diphoton~8 TeV ATLAS~\cite{Aad:2014ioa} and $7 + 8$~TeV CMS $WW$ and $ZZ$ searches~\cite{Khachatryan:2015cwa} are also shown for each channel.~In the gray shaded region we show for comparison the $s_\theta = 0.4$ (see~\eref{sth}) contour for single $H_5^0$ VBF production (see text).}
\label{fig:prodxbr}
\end{figure}
%

We see in~\fref{prodxbr} that by exploiting the $H^{0}_5 H^{\pm}_5$ Higgs pair production mechanism, custodial fiveplet scalars with masses $\lesssim 107$~GeV can be ruled out by 8 TeV diphoton searches, independently of the Higgs triplet vev.~We find similar limits as those found in~\fref{AAprodxbr} for the same values of suppressed and enhanced couplings to photons.~These are the first such limits on custodial fiveplet scalars and in particular, since the charged and neutral components are degenerate, limits from Tevatron $4\gamma + X$ searches~\cite{Aaltonen:2016fnw} do not apply.~This is because for cases like the custodial fiveplet where the masses are degenerate, the $H_5^{\pm} \to H_5^0 W^{\pm}$ decay is not available.~In this case the one loop $H_5^{\pm} \to W^{\pm} \gamma$ decay can become dominant leading instead to a $3\gamma + W$ signal.~Examining this decay as well should improve the sensitivity relative to diphoton searches.

To emphasize the utility of the DY pair production mechanism,~we also show (gray shaded region) the cross section times branching ratio assuming the VBF production mechanism.~Since for a fiveplet we have instead $\lambda_{WZ} = 1/2$ for the ratio of $WW$ and $ZZ$ couplings (see~\eref{lamWZ}), one cannot simply rescale the SM cross section for which $\lambda_{WZ} = 1$.~We therefore have again used~\cite{Alwall:2014hca,Hartling:2014xma} to obtain these results for 8 TeV LHC.~We have fixed $s_\theta = 0.4$ for the doublet-triplet vev mixing angle as defined in~\cite{Hartling:2014xma} and schematically in~\eref{sth}.~The value $s_\theta = 0.4$ is towards the upper limit of values still allowed by electroweak precision and 125~GeV Higgs data~\cite{Hartling:2014aga,Chiang:2015amq,Fabbrichesi:2016alj,Chiang:2016ydx}, but we can see in~\fref{prodxbr} this already renders diphoton searches for custodial fiveplet scalars based on VBF (and similarly for VH) production irrelevant.~We also emphasize that ruling out a custodial fiveplet below $\sim 110$~GeV independently of the vev
\footnote{In the case where the fermiophobic condition is relaxed and couplings to leptons are allowed, there is a previous vev-independent constraint on custodial-fiveplets from $pp \to H_5^{\pm\pm} H_5^{\mp\mp}$ and $pp \to H_5^{\pm\pm} H_5^{\mp}$ based on like-sign dimuon cross section limits~\cite{Kanemura:2014ipa} from ATLAS 8 TeV data~\cite{ATLAS:2014kca}.~This leads to a lower bound on the custodial fiveplet mass of about 76 GeV independently of the triplet vev~\cite{Logan:2015xpa}.~We thank Heather Logan for pointing this out.} 
allows us to unambiguously close the fiveplet `window' at masses below $\sim 100$~GeV~\cite{Logan:2015xpa} which is still allowed by electroweak precisions data~\cite{Englert:2013zpa} and essentially unconstrained by other LEP, Tevatron, and LHC direct searches.~Thus we are able to rule out an interesting region of parameter space of custodial Higgs triplet models which would otherwise be difficult to constrain directly.~We estimate 13 TeV diphoton searches will be sensitive to scalar masses up to $\sim 125$~GeV in the regime of dominant $W$ boson loop~\cite{Brooijmans:2016vro}, though NLO Higgs pair production effects~\cite{Degrande:2015xnm} may allow this to be extended
further.~The diphoton search discussed here may of course be useful for other scalars which are found in custodial Higgs triplet models, but we do not explore this here. 

Finally, we also see in~\fref{prodxbr} that $WW$ and $ZZ$ searches may be useful for probing custodial fiveplet scalars independently of the Higgs triplet vev as well.~Though 8 TeV searches are not quite sensitive, larger Higgs pair production cross sections at 13 TeV (see~\fref{HHprod}) should allow for fiveplet masses well above diphoton limits to be probed and possibly as high as $\sim 250$~GeV.~In particular, the $ZZ$ channel should become sensitive with early 13 TeV data for masses around the $ZZ$ threshold.~These also serves as a useful compliment to $W^+W^+$ searches for the doubly charged component of the custodial fiveplet~\cite{Englert:2013wga}.

\section{Summary and Conclusions}\label{sec:sum}

In this study we have explored diphoton and diboson searches at the LHC as probes of exotic fermiophobic Higgs bosons which are pair produced with an additional Higgs boson.~We have focused on the $pp \to W^{\pm} \to H^{\pm} H_F^0$ production channel which is present in all extensions of the SM Higgs sector and generally the dominant DY Higgs pair production mechanism.~We have emphasized that this production mechanism does not vanish in the limit of small exotic Higgs vacuum expectation value, unlike vector boson fusion and associated vector boson production.~Since measurements of the SM-like 125~GeV Higgs boson imply small exotic Higgs vacuum expectation values, previous searches for fermiophobic Higgs bosons which assumed vector boson fusion and associated vector boson production are now obsolete.

We have shown that by combining the Higgs pair production mechanism with diphoton searches, one can put stringent and rather generic bounds on fermiophobic Higgs bosons already with 8 TeV LHC diphoton data.~These limits are stronger and more general than those obtained in a very recent CDF $4\gamma + X$ search~\cite{Aaltonen:2016fnw} which are currently the only other relevant direct constraints on a fermiophobic Higgs boson in the small vev limit.~In particular, we find that while Tevatron diphoton searches are not sensitive to fermiophobic Higgs bosons, the larger Higgs pair production cross sections at 8 TeV LHC allow us to already generically rule out a neutral fermiophobic Higgs boson below $\sim 110$~GeV for degenerate masses and under the assumption that the couplings to photons are generated dominantly by a $W$ boson loop.~We have also emphasized that this degenerate mass scenario is not ruled out by Tevatron data.

If there is a mass splitting as large as $\sim 100$~GeV, we find masses below $\sim 100$~GeV can be excluded.~Furthermore, we find that if the couplings to photons are enhanced, masses up to $\sim 150$~GeV can be ruled out, while if cancellations conspire to give very small effective coupling to photons, fermiophobic Higgs scalars below $\sim 90$~GeV can still be ruled out by diphoton searches at 8 TeV.~This makes diphoton searches a robust and sensitive probe of lighter fermiophobic scalars.~Of course a dedicated multiphoton search at the LHC including the charged Higgs decay, as done at Tevatron, should improve limits further and is an important complementary probe.~We have also combined the $H^{\pm} H_F^0$ Higgs pair production channel with $WW$ and $ZZ$ diboson searches to probe fermiophobic Higgs masses up to $\sim 250$~GeV.~We find that while 8 TeV searches are not yet sensitive, the prospects for 13 TeV LHC are very promising if current limits can be improved by about an order of magnitude with future data.~The inclusion of NLO Higgs pair production effects as well as other subdominant production mechanisms may also further improve the limits discussed in this study.

Finally, we have examined the particular case of a custodial fiveplet scalar found in all incarnations of custodial Higgs triplet models~\cite{Georgi:1985nv,Hartling:2014zca,Cort:2013foa} in which the neutral and charged component are predicted to be degenerate.~Thus the CDF $4\gamma + X$ search~\cite{Aaltonen:2016fnw} cannot be applied to this case.~We have shown for the first time that a custodial fiveplet scalar below $\sim 110$~GeV is ruled out by 8 TeV diphoton searches and possibly up to higher masses if charged scalar loops produce large constructive contributions to the effective photon couplings.~These limits are also largely independent of the Higgs triplet vev and so robustly close the `fiveplet window' at masses below $\sim 110$~GeV~\cite{Logan:2015xpa}, still allowed by electroweak precision and 125~GeV Higgs boson data.~We also find that diboson searches, and in particular $ZZ$ searches, may be useful for larger fiveplet masses, allowing us to potentially obtain limits again independently of the Higgs triplet vev.
 
To summarize, by combining the $pp \to W^{\pm} \to H^{\pm} H_F^0$ Higgs pair production mechanism with $H_F^0 \to VV$ diphoton and diboson decays, one obtains a powerful probe at the LHC of fermiophobic Higgs bosons for masses up to $\sim 250$~GeV.~These searches are sensitive even in the limit of vanishing exotic Higgs vev and open a yet to be explored avenue to search for fermiophobic Higgs bosons at the LHC in both current and future data.

\bigskip

{\em Acknowledgments:} 
We would like to thank Yi Chen,~Jorge de Blas, Francisco del Aguila,~Adam Falkowski,~Heather Logan,~Javi Serra,~Daniel Stolarski,~and Roberto Vega for helpful discussions.~R.V.M. would like to especially thank Kunal Kumar for help with Madgraph/Feynrules model implementations.~We also would like to thank the Ecole de Physique des Houches for creating a stimulating atmosphere where this project was started.~The work of A.D. is partly supported by the National Science Foundation under grant PHY-1520966. The work of J.S. and R.V.M.~is supported by MINECO, under grant number FPA2013-47836-C3-2-P.~J.S.~is also supported by the European Commission contract PITN-GA-2012-316704 (HIGGSTOOLS) and by Junta de Andaluc\'{\i}a grants FQM 101 and FQM 6552.~The work of M.G. P. and M.Q.~is partly supported by MINECO under Grant CICYT-FEDER-FPA2014-55613-P, by the Severo Ochoa Excellence Program of MINECO under Grant SO-2012-0234, and by Secretaria d'Universitats i Recerca del Departament d'Economia i Coneixement de la Generalitat de Catalunya under Grant 2014 SGR 1450.


\bibliographystyle{apsrev}
\bibliography{refs}

\end{document}